\documentclass[useAMS]{mn2e}
\usepackage{graphicx}
\usepackage{longtable}

\def\oversim#1#2{\lower0.5pt\vbox{\baselineskip0pt \lineskip-0.5pt
     \ialign{$\mathsurround0pt #1\hfil##\hfil$\crcr#2\crcr\sim\crcr}}}
\def\lsim{\mathrel{\mathpalette\oversim<}}    

\title[AGB Variables and the Mira Period-Luminosity Relation]
{AGB Variables and the Mira Period-Luminosity Relation}
\author[Whitelock et al.]{Patricia A. Whitelock$^{1,2,3}$, Michael W. Feast$^2$
and Floor van Leeuwen$^{4}$\\
      $^1$ South African Astronomical Observatory, P.O.Box 9, 7935
           Observatory, South Africa (email: paw@saao.ac.za).\\
      $^2$ Astronomy Department, University of Cape Town, 7701 Rondebosch,
           South Africa.\\
      $^3$ National Astrophysics and Space Science Programme, Department of
           Mathematics and Applied Mathematics,\\ 
           University of Cape Town, 7701 Rondebosch, South Africa.\\
$^4$ Institute of Astronomy, Madingley Rd, Cambridge, England.\\
}

\begin{document}
\maketitle

\begin{abstract} 
 Published data for large amplitude asymptotic giant branch variables in
the Large Magellanic Cloud are re-analysed to establish the constants for an
infrared ($K$) period-luminosity relation of the form: $M_K=\rho[\log P-2.38]
+ \delta$. A slope of $\rho=-3.51\pm0.20$ and a zero point of
$\delta=-7.15\pm0.06$ are found for oxygen-rich Miras (if a distance modulus 
of $18.39\pm0.05$ is used for the LMC). Assuming this slope is applicable to
Galactic Miras we discuss the zero-point for these stars using the revised
{\it Hipparcos} parallaxes together with published VLBI parallaxes for OH
Masers and Miras in Globular Clusters. These result in a mean zero-point of
$\delta=-7.25\pm0.07$ for O-rich Galactic Miras.  The zero-point for Miras
in the Galactic Bulge is not significantly different from this value.

Carbon-rich stars are also discussed and provide results that are consistent
with the above numbers, but with higher uncertainties. Within the
uncertainties there is no evidence for a significant difference between the
period-luminosity relation zero-points for systems with different
metallicity.
\end{abstract}

\begin{keywords}{stars:AGB and post-AGB -- stars: oscillations -- stars: 
carbon -- stars: variables: other} 
\end{keywords}

\section{Introduction}
 Large amplitude Asymptotic Giant Branch (AGB) variables (Miras) are
important distance indicators for old and intermediate age populations. They
are luminous, both bolometrically and in the near-infrared, and easily
identified by their late spectral types (Me, Ce, Se or very rarely Ke),
large amplitudes ($\Delta V > 2.5$ mag, $\Delta K > 0.4$ mag) and long
periods ($100 \lsim P \lsim 1000$). The increasing use of adaptive optics on
large telescopes at near-infrared wavelengths to study stellar populations
(e.g. Da Costa 2004) at large distances will require confidence in the
calibration of the AGB variables as distance indicators. Alternatively, if
the distance is known the luminosities of the brightest AGB stars provide
insight into the intermediate age populations, e.g. Menzies et al. (2008).

 Wood et al. (1999) demonstrated that, within the LMC, AGB variables fall on
a series of, approximately parallel, period-luminosity (PL) relations at
$K$, (see also Cioni et al. 2001, Ita et al. 2004, Fraser et al. 2005, 
Soszy\`nski et al. 2007). The large amplitude variables, i.e. the Miras,
however, lie only on a singe PL($K$) relation. The existence of a Mira
PL($K$) relation has been known for some while (Feast et al. 1989; Hughes \&
Wood 1990) and is now generally thought to represent the relation for
fundamental pulsation. While some of the low amplitude variables also
pulsate in the fundamental, others show various overtones. Low amplitude
variables are of limited use for distance scale studies as it is not simple
to establish upon which of the various PL relations they lie.  Complications
do arise at periods in excess of about 400 d where some Miras in the LMC
have higher luminosities, possibly as a consequence of hot bottom burning
(e.g. Whitelock et al. 2003). Feast (2004) provides a recent discussion of
AGB stars as distance indicators and of the zero-point of the PL relation.

 Following a brief re-examination of the Mira PL($K$) relation in the LMC, we
take advantage of the new analysis of the {\it Hipparcos} data (van Leeuwen
2007a and 2007b, see also van Leeuwen 2005 and van Leeuwen \& Fantino 2005) to
re-examine the distance scale for large amplitude AGB variables within the
Galaxy. Analysis of the original {\it Hipparcos} data was discussed by van
Leeuwen et al. (1997), Whitelock \& Feast (2000 - henceforth Paper~I) and
Knapp et al. (2003).

In the last part of this paper we put together all the available information
on Mira parallaxes to establish the best value for the zero-point of the
Mira PL($K$) relation for the Galaxy and compare this with values for
elsewhere.

\begin{table}
\centering
\caption{LMC variables used to establish the slope for the Mira
PL($K$) relation}\label{lmcdata}
\begin{center}
\begin{tabular}{lccc}
\hline
\multicolumn{1}{c}{name}& P & $K$ & note\\
& \multicolumn{1}{c}{(d)} &  \multicolumn{1}{c}{(mag)} \\
\hline
\multicolumn{4}{c}{O-rich stars}\\
0517--6551     & 116 & 12.25  & \\
C38            & 130 & 12.12  & \\ 
0512--6559     & 141 & 12.13  & \\ 
W132           & 156 & 11.67  & \\ 
0526--6754     & 157 & 11.79  & \\ 
W151           & 174 & 11.74  & \\ 
W148           & 183 & 11.82  & \\ 
W158           & 194 & 11.77  & \\ 
0528--6531     & 195 & 11.48  & \\ 
C11            & 202 & 11.51  & \\ 
GR13           & 202 & 11.59  & \\ 
SHV05220--7012 & 205 & 11.73  & 1\\
0507--6639     & 208 & 11.57  & \\ 
C20            & 210 & 11.54  & \\ 
W77            & 213 & 11.25  & S\\
R120           & 217 & 11.38  & \\ 
W94            & 220 & 11.28  & \\ 
W74            & 231 & 11.49  & \\ 
WBP74          & 233 & 11.50  & 1\\
W1             & 235 & 11.48  & \\ 
W140           & 243 & 11.19  & \\ 
0533--6807     & 247 & 11.38  & \\ 
R141           & 258 & 10.99  & \\ 
R110           & 261 & 11.29  & \\ 
W48            & 279 & 10.99  & \\ 
0537--6607     & 284 & 11.02  & \\ 
0505--6657     & 307 & 10.67  & \\ 
0524--6543     & 315 & 10.71  & \\ 
W126           & 318 & 10.89  & K\\ 
SHV05305--7022 & 362 & 10.57  & \\ 
R105           & 413 & 10.33  & \\ 
\multicolumn{4}{c}{C-rich stars}\\
0530--6437     & 157 & 12.08  & \\ 
0515--6617     & 226 & 11.16  & \\ 
0528--6520     & 229 & 11.08  & \\ 
0520--6528     & 233 & 11.28  & \\ 
0519--6454     & 242 & 11.09  & \\ 
W220           & 281 & 10.83  & \\ 
0529--6759     & 283 & 10.91  & \\ 
0515--6451     & 284 & 10.81  & \\ 
SHV05027--6924 & 298 & 10.82  & \\
0514--6605     & 308 & 10.64  & \\ 
0534--6531     & 308 & 10.98  & \\ 
0529--6739     & 319 & 10.60  & \\ 
0502--6711     & 322 & 10.53  & \\ 
C7             & 327 & 10.69  & \\
0541--6631     & 342 & 10.50  & \\ 
R153           & 347 & 10.52  & \\ 
WBP14          & 351 & 10.62  & \\
W103           & 363 & 10.78  & \\ 
0515--6438     & 365 & 10.90  & \\ 
0537--6740     & 367 & 10.47  & \\ 
SHV05003--6817 & 369 & 10.58  & \\
SHV05260--7011 & 373 & 10.54  & \\
\hline
\multicolumn{4}{l}{notes 1. Period taken from MACHO;}\\
\multicolumn{4}{l}{S and K are spectral types}\\
\hline
\end{tabular}
\end{center}
\end{table}

\begin{figure*}
\includegraphics[width=16cm]{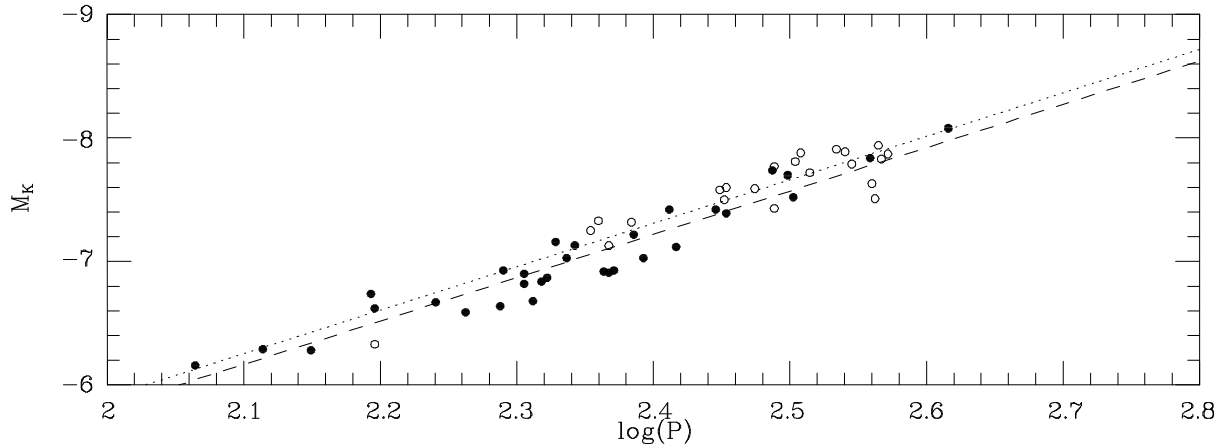}
\caption{The PL($K$) relation for Mira variables in the LMC. Solid and open 
symbols are O- and C-rich respectively. The dashed line is the fit to the
O-rich stars from Table~\ref{lmcpl} assuming a distance modulus for the LMC
of 18.39 mag, while the dotted line is the relation for C-stars.}
\label{fig_lmcpl}
\end{figure*}

\section{The LMC PL($K$) relation}
 Table~\ref{lmcdata} lists the data for the LMC stars, with periods less
than 420 d, which we use here to establish the PL($K$) relation. The
starting point is the material from Feast et al. (1989) for large amplitude
variables with multiple observations. This is modified in light of Glass \&
Lloyd Evans (2003) who used MACHO data to refine the periods for the same
group of stars and to eliminate three of them (W19, W30, W46) as
semi-regular (SR), rather than Mira, variables. To this is added the
observations from Whitelock et al. (2003) for three new O-rich and four
C-rich Miras within our period range, as well as additional material for two
O-rich stars (R105, 0517--6551) already in the sample. The changes
introduced by the two 2003 papers are minor and this data-set is only
marginally different from that used by Feast et al. (1989).
 
Twenty nine M-type Miras together with one S-type and one K-type Mira are
analysed together as O-rich stars. An interstellar extinction correction of
$A_K=0.02$ mag (Feast et al. 1989) was applied to the data in
Table~\ref{lmcdata}. The PL($K$) relation is fitted in the form:
\begin{equation} 
M_K =  \rho [\log P-2.38] + \delta.
\end{equation}
 Where $M_K$ is the absolute $K$ mag, $P$ the pulsation period of the Mira
and $\delta$ the zero-point of the PL($K$) relation, which has a slope $\rho$.
In previous work, by ourselves and others, zero-points have been derived at
$\log P=0$, far outside the range of Mira periods. The distance modulus of
the LMC is taken to be $(m-M)_{LMC}=18.39\pm0.05$ (van Leeuwen et al. 2007)


Table~\ref{lmcpl} lists the values of the slope and zero-point found by
least squares fitting separately to the O-rich stars and to the C-rich stars
as well as to both groups together; $\sigma$ is the standard deveation of
the best fit. The results are illustrated in
Fig.~\ref{fig_lmcpl}. There is no significant difference in the slope
determined for the O- and C-rich Miras. If we use the slope found for the
O-rich sample and apply it to the C-rich sample we find a zero point of
$\delta+(m-M)_{LMC}=11.148\pm0.032$. Thus the difference between the zero-points
for the C- and O-rich Miras is $0.093\pm0.041$.

The slope determined above for the LMC O-rich Miras, i.e. $\rho=-3.51\pm
0.20$ can be compared with the value of $-3.59\pm0.06$ obtained by Ita et
al. (2004) for LMC stars close to PL($K$) sequence C and selected by colour
to be O-rich. While the error shows this number is well defined, Ita et al.
include some low amplitude, i.e. non-Mira, variables in the fit.
Furthermore, it seems tautological to define the PL relation in terms of
stars that were selected because they fell on sequence C.  Rejkuba (2004)
found a very similar slope, $-3.37\pm0.11$, for colour selected Miras in
Cen~A, supporting the assumption that this PL($K$) relation is universal.

\begin{table*}
\centering
\caption{Solutions of Equation 1}\label{lmcpl}
\begin{center}
\begin{tabular}{cccccc}
\hline
\multicolumn{1}{c}{typ}& \multicolumn{1}{c}{No.}&
\multicolumn{1}{c}{$\sigma$}& 
 \multicolumn{1}{c}{$\rho$}& \multicolumn{1}{c}{$\delta+(m-M)_{LMC}$} & $\delta$\\
\hline
O- & 31 & 0.14 & $-3.51\pm 0.20$& $11.241\pm0.026$ & $-7.15\pm0.06$\\
C- & 22 & 0.15 & $-3.52\pm 0.36$& $11.149\pm0.047$ & $-7.24\pm0.07$\\
all &    53 & 0.15 & $-3.69\pm 0.16$ & $11.206\pm0.020$& $-7.18\pm0.05$\\
\hline
\end{tabular}
\end{center}
\end{table*}

\section{The Revised {\it Hipparcos} Sample and Associated Data}
 Of the sample selected in Paper~I there are astrometric data in the revised
{\it Hipparcos} catalogue (van Leeuwen 2007) for 184 O-rich Miras, 15 O-rich
semi-regulars and 40 C-rich variables.

 In Paper~I a number of non-Mira variables were included in the selection on
the basis of their spectra, which were Mira-like in that they showed the
emission lines that are characteristic of the shock waves associated with
pulsations in Miras. For the O-rich stars 4 of the 15 semi-regular variables
included in the sample have large amplitudes ($\Delta H_p>1.5$ mag, where
$H_p$ is the broadband visual mag measured by {\it Hipparcos}) and it is
probably only these 4, T~Ari, T~Cen, W~Hya and TV~And, that we should
include with the Miras in the parallax analysis. For the C-rich stars it is
actually very difficult to distinguish between Miras and non Miras (see also
Whitelock et al. 2006).

 The infrared and associated data used in the following analysis are
listed in the Appendix (Table~\ref{dist}), which also includes all details
which differ from Paper~I.

\subsection{The SP-red and SP-blue stars}
 Whitelock et al. (2000) divided the stars with periods below 225 d into two
groups on the basis of their $Hp-K$ colour, and called them the short period
red (SP-red) and short period blue (SP-blue) groups. The analysis of Paper~I
suggested that the SP-reds were more luminous, at a given period, than 
the SP-blues. Most critically a kinematic analysis (Feast \& Whitelock 2000)
indicated that the SP-reds were younger than the SP-blues which were more
akin to the Globular Cluster Miras and a natural extrapolation of the Miras
with $P>225$ d to shorter periods. 

\section{Analysis of {\it Hipparcos} Data}
 We follow the same procedure as in Paper~I and the details are not repeated
here, but it is useful to give the formulae which differ slightly due to the
reformulation of the PL($K$) relation described in section 2. We assume
throughout that the the slope derived above from the O-rich LMC Miras
will be the same for the Galaxy.  Thus equation 1 with, $\rho=-3.51\pm0.20$, 
is solved for $\delta$, as previously, in the form:
\begin{equation}
10^{0.2 \delta}= 0.01 \pi 10^{0.2(3.51[\log P-2.38] +K_0)},
\end{equation}
where $\pi$ is the parallax in milliarcsec (mas) and $K_0$ is the mean $K$
mag corrected for interstellar extinction (see Appendix). The right hand
side of equation 2 is weighted by the following expression:
\begin{equation} 
{\rm weight} = 1/[\sigma 10^{0.2(K_0+3.51[\log P-2.38])-2.0}]^2,
\end{equation}
where
\begin{equation}
\sigma^2=\sigma^2_\pi+(0.4605)^2\pi^2_{PL(K)}[\sigma^2_K+\sigma^2_{PL(K)}],
\end{equation}
with $\sigma_\pi$ the error on the parallax as quoted in the revised {\it
Hipparcos} catalogue, $\pi_{PL(K)}$ the photometric parallax derived from
the PL($K$) relation, $\sigma_{PL(K)}$ the standard deviation from the PL($K$)
relation (0.14 and 0.15 for the O- and C-rich stars respectively,
see Table~\ref{lmcpl}), and $\sigma_K$ is the uncertainty associated with
individual $K_0$ mags. The latter term is evaluated as follows: $\sigma_K
=0.3 \Delta K/\sqrt{N}$ (where $N$ observations were used to derive the mean
$K_0$). For further details refer to Paper~I.

Due to  uncertainties in the adopted slope, the error in a predicted
absolute magnitude at any $\log P$ is, in the case of the O-Mira solution of
Table~\ref{lmcpl}:\begin{equation}
\sqrt ([(\log P -2.38)0.20]^{2} + [0.06]^{2}),
\end{equation}
and similar relations apply in other cases.

R Leo has a parallax measured by the Allegheny Observatory at $\pi = 8.3
\pm 1.0$ mas (Gatewood 1992). For the analysis this is combined with the
{\it Hipparcos} value, $\pi = 14.03 \pm 2.65$, to give $\pi = 9.01 \pm
1.42$.

\subsection{Zero-point from the {\it Hipparcos} parallaxes}
\subsubsection{O-rich stars}
 The first part of Table~\ref{zp} lists various values of $\delta$, from
equation (2) [weighted according to equations (3) and (4)], derived from
different subgroups of the {\it Hipparcos} data. In examining these results
it is crucial to remember that most of the weight resides with a small
number of stars. 

Solution 1 shows the result of using all of the O-rich stars for
completeness and comparison with Paper~I, although it is quite clear that
this is not a useful solution as there are stars included in the full group
that are not large amplitude variables and which certainly lie on one of the
PL($K$)s relations above the one for Miras (see also Glass \& van Leeuwen
2007). W Cyg is the most obvious example and solution 2 shows the effect of
leaving it out. In fact solution 2 is very close to the best solution we
obtain.

Solutions 3 and 4 are for large amplitude variables and Mira variables
respectively; the results are very similar as might be expected.

Solutions 5 to 8 show the results of separating out the SP-red and SP-blue
stars (see Section 3.1). In fact there is no evidence here for a significant
difference between the groups, provided that W~Cyg is omitted, with the
revised {\it Hipparcos} data.

Solution 9 omits the small amplitude variables and the SP-red stars and can
be compared with $\beta=0.84 \pm 0.14$ ($\delta\sim-7.51$ for a slightly
different slope for the PL($K$) relation), the solution of choice from
Paper~I. Solution 10 rejects the low amplitude non-Miras and takes only the
42 stars with $\rm weight >1.3\times10^3$. If we were to decide that the
SP-reds should be included, despite their kinematic differences from the
SP-blues, then this solution, with its marginally smaller error, might be
preferred, but it makes very little practical difference.

Solutions 11 and 12 show that much the same value of $\delta$ is derived
from the longer period stars alone. It is important to note there is thus no
reason to suspect that they are brighter or fainter than their short period
counterparts (see also sections 5.1.1 and 5.1.2).

Solution 13 was derived in a different way and is discussed below
(section 4.1.3).

\subsubsection{C-rich stars}
 Solutions 14 to 24 apply to subsets of the data for C-rich stars. We note
again that it is difficult to distinguish between Mira and non-Mira C-stars
and that the amplitudes of the C-rich stars are on average distinctly lower
than those of their O-rich counterparts.  As in Paper~I we note that WZ~Cas
and WX~Cyg are lithium rich and therefore likely to differ from other stars
in the sample. WZ~Cas has a high weight and is clearly more luminous than
the bulk of the sample. WX~Cyg makes little difference, but we leave both
stars out for most of the solutions (17 to 24).

 The errors are high and it is difficult to deduce much about the
sub-groupings of C stars. The non-Miras may have a lower value of $\delta$
(solution 23) than the Miras (solution 17) or the large amplitude variables
(solution 18); this is to be expected if some of the low amplitude variables
lie on overtone sequences, but cannot be claimed with confidence. Solution
24 is probably the best, including only the higher weight stars with large
amplitudes. V~Hya is in a binary system (see Olivier, Whitelock \& Marang
(2001) and references therein), although its normal variations are due to
pulsation. Solution 19 was derived leaving out V~Hya.

 These various solutions indicate that within the errors the C- and O-rich
stars obey the same PL relation at $K$, which is consistent with earlier
findings from the LMC (section 2 and Feast et al. 1989). 

\begin{table*}
\centering
\begin{minipage}{110mm}
\caption{PL($K$) Zero-Point ($\delta$)}\label{zp}
\begin{center}
\begin{tabular}{@{}lrrrcl}
\hline
\multicolumn{1}{c}{Soln.} &
\multicolumn{1}{c}{No.} & \multicolumn{1}{c}{weight} & 
\multicolumn{1}{c}{$\delta$} & \multicolumn{1}{c}{$\sigma_{\delta}$} & 
\multicolumn{1}{l}{stars included in the analysis} \\
\multicolumn{1}{c}{No.} & \multicolumn{1}{c}{stars} 
& \multicolumn{1}{c}{$\times 10^{-3}$} & (mag)\\
\hline
& & \multicolumn{4}{l}{Oxygen-rich stars from {\it Hipparcos}}\\
$\,\ $1 & 199 & 710 & --7.46 & 0.11 & all \\
$\,\ $2 & 198 & 572 & --7.24 & 0.11 & all but W Cyg\\
$\,\ $3 & 182 & 453 & --7.27 & 0.13 & $\Delta Hp > 1.5$ mag\\
$\,\ $4 & 184 & 385 & --7.22 & 0.14 & Mira variables\\
$\,\ $5 & 183 & 460 & --7.28 & 0.13 & all but SP-red stars\\
$\,\ $6 &  37 &   & --7.59 & 0.46 & SP-blue stars\\
$\,\ $7 & 16 & 259 & --7.81 & 0.22 & SP-red stars\\
$\,\ $8 & 15 & 112 & --7.11 & 0.14 & SP-red stars not W Cyg\\ 
$\,\ ${\bf 9} & {\bf 168} & {\bf 434} & {\bf --7.27} & {\bf 0.14} & {\bf not SP-red stars; $\Delta Hp > 1.5$ mag}\\
{\bf 10} & {\bf 42} & {\bf 415} & {\bf--7.32} & {\bf 0.10} & {\bf Miras; non-Miras with $\Delta Hp > 1.5$
mag; weight$>1.3\times10^3$}\\
11 &146 & 433 & --7.26 & 0.14 & $\rm P > 224 $ day\\
12 & 21&  84 & --7.04 & 0.27 & $\rm P \ge 400$ Miras \& non-Miras with $\Delta Hp > 1.5$ mag; \\
\\
13 &  6 &     &--7.11 & 0.17 & $\sigma_\pi/\pi<0.16$, Miras, non-Miras with $\Delta Hp > 1.5$ mag; \\
\\
\hline
& \multicolumn{5}{l}{Carbon-rich stars from {\it Hipparcos}} \\
14 & 40 & 97 & --7.91 & 0.41 & all \\
15 & 39 & 72 & --7.55 & 0.40 & omitting WZ Cas\\
16 & 38 & 72 & --7.55 & 0.40 & omitting WZ Cas and WX Cyg\\
& \multicolumn{5}{l}{Carbon-rich stars omitting WZ Cas and WX Cyg} \\
17 & 23 & 24 & --7.21 & 0.65 & Miras only\\
18 & 31 & 33 & --7.14 & 0.50 & $\Delta Hp>1.0$ mag\\
19 & 30 & 30 & --7.03 & 0.50 & $\Delta Hp>1.0$ mag, omitting V Hya\\
20 & 24 & 46 & --7.74 & 0.60 & $\rm P < 400$\\ 
21 & 14 & 26 & --7.24 & 0.50 & $\rm P \ge 400$\\ 
22 & 35 & 70 & --7.57 & 0.43 & $Hp-K < 8.5$\\ 
23 & 15 & 48 & --7.73 & 0.45 & non Miras\\
24 & 16 & 32 & --7.18 & 0.37 & $\Delta Hp>1.0$ mag, $\rm weight>0.24\times10^3$\\
\\
\hline
& \multicolumn{5}{l}{Values derived in other ways} \\
25 & 5 & & --7.08 & 0.17 & VLBI parallaxes for OH-Miras (section 5.1.1,
Table~\ref{vlbi})\\
26 & 11 & & --7.34 & 0.13 & Globular Cluster Miras (section 5.2)\\
27 & 31 & & --7.15 & 0.06 & LMC O-rich Miras (section 2)\\
28 & 22 & & --7.24 & 0.07 & LMC C-rich Miras (section 2)\\
29 & 54 & & --7.04 & 0.11 & Miras in the Galactic Bulge (section 5.3)\\
\hline
\end{tabular}
\end{center}
\end{minipage}
\end{table*}

\subsubsection{Individual stars with low uncertainty}
 As an alternative approach, because there are a few stars with good
signal-to-noise ratios, we selected stars with positive {\it Hipparcos}
parallaxes for which $\sigma_\pi/\pi<0.16$ (Table~\ref{indiv}). These are
the same stars which dominate the various solutions discussed in section
4.1.1, as the 8 stars involved have 62 percent of the weight of the complete
sample of 199 O-rich stars. There are no C-stars with $\sigma_\pi/\pi<0.16$.
The 8 include 3 non-Mira variables, W~Cyg, L2~Pup and W~Hya, of which only
the latter is a large amplitude variable. In view of the fact that these
stars were selected on the basis of their $\sigma_{\pi}/\pi$ ratio a
Lutz-Kelker correction has been applied to their apparent magnitudes. This
is evaluated (on the model used by Benedict et al. (2007)) as:
$$LK=-8.09(\sigma_\pi/\pi)^2,$$ and it makes them up to 0.2 mag
brighter than they would otherwise be. The corrections are listed in
Table~\ref{indiv}.

\begin{figure*}
\includegraphics[width=16cm]{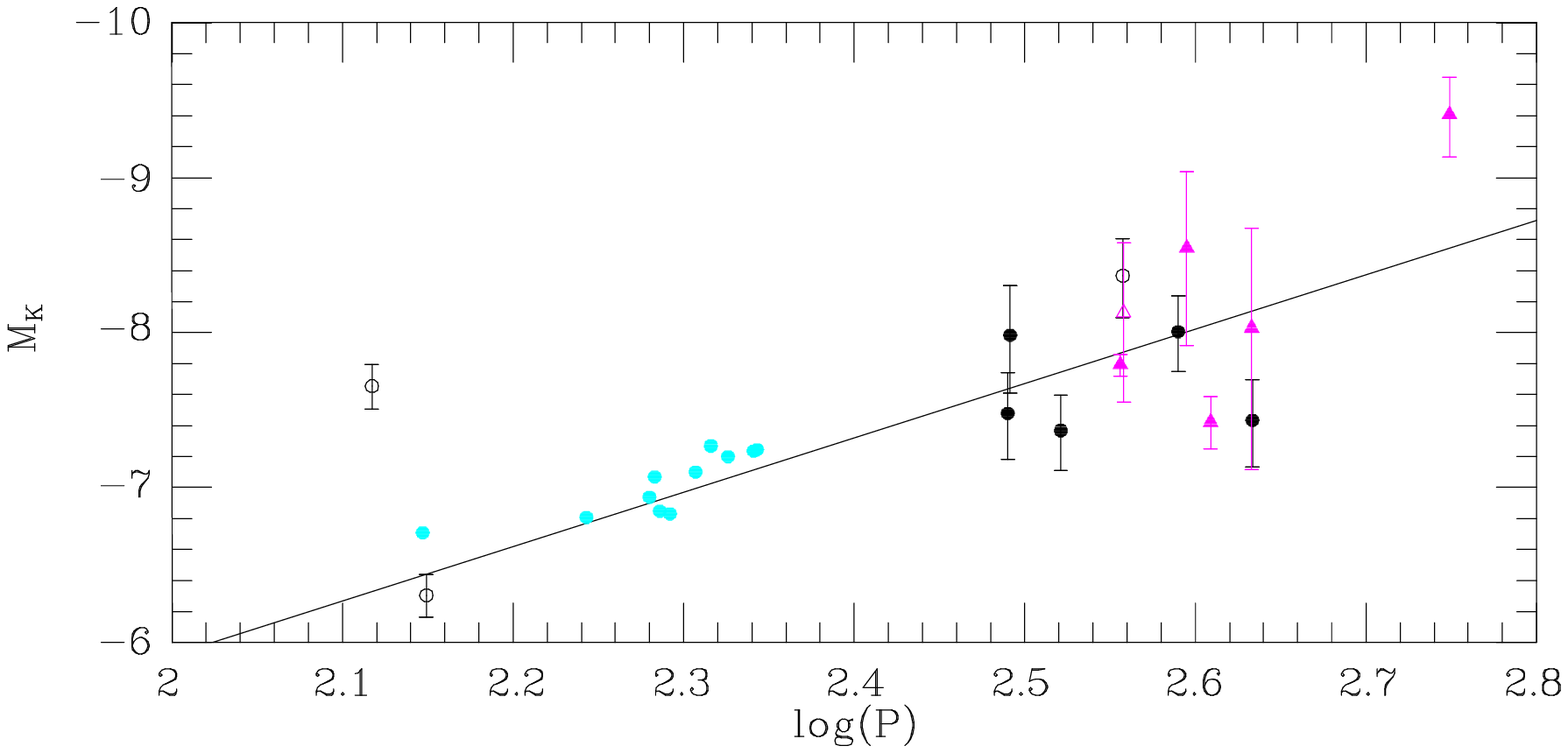}
\caption{The PL($K$) relation for AGB variables with good individual distances. 
The closed and open symbols represent Miras and SR variables respectively;
the points without error bars are for the globular cluster Miras, the circles
with errors bars represent the {\it Hipparcos} data with high S/N (Table~4)
and the triangles represent VLBI parallaxes (Table~5). The line is for
$M_K=-3.51 [\log P-2.38] -7.25$.}
\label{fig_pl}
\end{figure*}

 These stars are shown in the PL($K$) diagram, Fig.~\ref{fig_pl}, where they
are compared with the stars discussed below. The PL($K$) relation zero point
derived from the six stars W Hya, R Hya, R Cas, R Car, $o$ Cet and R~Leo,
$\delta=-7.11\pm0.17$, is listed as solution 13 in Table~\ref{zp}. Note that
the two low amplitude semi-regulars, W Cyg and L$_2$ Pup, which have
$\sigma_\pi/\pi<0.16$ and are shown in the figure are omitted from this
solution for consistency. 

\begin{table*}
\centering
\caption{Data for individual stars with $\sigma_\pi/\pi<0.16$}\label{indiv}
\begin{tabular}{rcrcccc}
\hline
\multicolumn{1}{c}{name} & 
\multicolumn{1}{c}{$\pi$} & \multicolumn{1}{c}{$K_0$} & 
\multicolumn{1}{c}{$LK$} &type & $\log P$
& $M_K$ \\
\hline
$o$ Cet & $ 10.91\pm 1.22$ & --2.45 & --0.10 & M  & 2.521 & $-7.37^{+.26}_{-.23}$ \\
L$_2$ Pup  & $ 15.61\pm 0.99$ & --2.24 & --0.03 &SR & 2.149 & $-6.31^{+.14}_{-.13}$ \\
R Car   & $  6.34\pm 0.81$ & --1.35 & --0.13 & M  & 2.490 & $-7.48^{+.30}_{-.26}$ \\
R Leo   & $  9.01\pm 1.42$ & --2.55 & --0.20 & M  & 2.491 & $-7.98^{+.37}_{-.32}$ \\
R Hya   & $  8.24\pm 0.92$ & --2.47 & --0.10 & M  & 2.590 & $-8.01^{+.26}_{-.23}$ \\
W Hya   & $  9.59\pm 1.12$ & --3.16 & --0.11 &SR & 2.558 & $-8.37^{+.27}_{-.24}$ \\
W Cyg   & $  5.72\pm 0.38$ & --1.40 & --0.04 & SR & 2.117 & $-7.65^{+.15}_{-.14}$ \\
R Cas   & $  7.95\pm 1.03$ & --1.79 & --0.14 &M  & 2.633 & $-7.43^{+.30}_{-.26}$ \\
\hline
\end{tabular}
\end{table*}

\section{Other Miras with Accurate Distances}
 
\subsection{Miras with VLBI parallaxes}
 Parallaxes for the OH and H$_2$O masers associated with Miras promise
to provide accurate distances for significant numbers of AGB
stars. These techniques will be particularly important in establishing the
distances to stars with high mass-loss rates, for which circumstellar
extinction makes optical measurements extremely difficult.

\subsubsection{Miras with OH parallaxes}
 Vlemmings et al. (2003) and Vlemmings \& van Langevelde (2007) have
published the results of astrometry for 5 stars, which are listed in
Table~\ref{vlbi} together with mean $K$ mags from SAAO photometry or the
literature after correcting for interstellar extinctions determined as
described in the Appendix. The {\it Hipparcos} and VLBI parallaxes agree
within the quoted uncertainties.

 The parallaxes of these 5 OH Masers, alone, give a zero-point for the PL($K$)
relation of $\delta=-7.08\pm0.17$. As Vlemmings \& van Langevelde (2007)
pointed out, the results for U~Her seem to be different from the others, but
taken with the {\it Hipparcos} results the difference is not obviously
significant.

 van Langevelde, van der Heiden \& van Schooneveld (1990) measured phase-lag
distances to more than 12 OH/IR sources (essentially Miras with high
mass-loss rates and therefore thick dust shells). The $K$ magnitudes of
these stars experience significant circumstellar extinction and they cannot
therefore easily be presented in a PL($K$) relation. Their position in a
PL($M_{bol})$ relation was discussed by Whitelock, Feast \& Catchpole (1991
their fig.~10) from which they appear to be consistent with the
PL($M_{bol}$) relation extrapolated from shorter periods. Phase-lag
distances might still prove useful for the longest period objects although
it will be difficult to bring the errors down so that they can compete with
VLBI parallaxes.

\begin{table*}
\centering
\caption{Data for stars with VLBI parallaxes}\label{vlbi}
\begin{tabular}{rrcrcccc}
\hline
\multicolumn{1}{c}{name} & \multicolumn{1}{c}{$\pi_{VLBI}$} & 
\multicolumn{1}{c}{$\pi_{Hip}$} & \multicolumn{1}{c}{$K_0$} & typ & $\log P$
& $M_K$ & ref \\
\hline
S CrB  & $2.39\pm 0.17$  &  $1.85\pm 1.19$ &  0.32& M& 2.556&$-7.79^{+.07}_{-.07}$ &V07\\
U Her  & $3.76\pm 0.27$  &  $4.06\pm 1.19$ &--0.30& M& 2.609&$-7.42^{+.17}_{-.17}$ &V07\\
RR Aql & $1.58\pm 0.40$  &  -              &  0.46& M& 2.595&$-8.55^{+.63}_{-.49}$ &V07 \\
 W Hya &$10.18\pm 2.36$  &  $9.59\pm 1.12$ &--3.17&SR& 2.558&$-8.13^{+.57}_{-.45}$ &V03\\
 R Cas &$ 5.67\pm 1.95$  &  $7.95\pm 1.03$ &--1.80& M& 2.663&$-8.03^{+.92}_{-.64}$ &V03\\
UX Cyg & $0.54\pm 0.06$ &  -               &  1.93& M& 2.749&$-9.41^{+.27}_{-.24}$ &K05\\
\hline
\multicolumn{6}{l}{refs: V07 Vlemmings \& van Langevelde (2007);}\\  
\multicolumn{6}{l}{V03 Vlemmings et al. (2003);}\\ 
\multicolumn{6}{l}{K05 Kurayama et al. (2005).}
\end{tabular}
\end{table*}

\subsubsection{Mira with H$_2$O parallax}
 Kurayama, Sasao \& Kobayashi (2005) published an accurate H$_2$O maser
parallax for UX~Cyg which is given in Table~\ref{vlbi} and illustrated in
Fig.~\ref{fig_pl}. It lies distinctly above the Mira PL($K$) relation. This
may indicate that it is an overtone pulsator and that it therefore lies on
one of the other period-luminosity relations discussed by Wood (2000),
although it has a longer period than any of the LMC stars on those
sequences. The stars that define the other PL($K$) relations are low
amplitude semi-regular variables, and it would be surprising to find a star
like UX~Cyg in this position. However, the PL($K$) sequences were defined
for stars in the Magellanic Clouds, all of which are at the same distance,
and very little is actually know about the distances of stars with periods
significantly longer than 400 days in the Galaxy (except close the the
Galactic Centre, where interstellar extinction is very high and patchy).
Note that solution 12 (Table~\ref{zp}) for the 21 {\it Hipparcos} stars with
$\rm P \ge 400$ days gives a zero-point completely consistent with them
lying on the same PL($K$) relation as the shorter period objects.

 The alternative explanation is that UX~Cyg is more luminous than those on
the PL($K$) relation as a result of hot bottom burning (HBB), and is
therefore akin to the luminous Magellanic Cloud Miras discussed by Whitelock
et al. (2003). This could be confirmed by the detection of abundance
anomalies associate with HBB, which include a measurable lithium content.
Note that the long period O-rich LMC Miras that Whitelock et al. (2003)
thought lay close to the extrapolated {\it bolometric} PL relation, would
actually lie below a PL($K$) relation because the significant circumstellar
reddening would effect their apparent $K$ luminosities.

UX Cyg is not included in evaluating the mean PL($K$) relation zero-point
below.

\subsection{Globular Clusters}
 Feast et al. (2002) discussed the PL relation for globular cluster Miras,
using a distance scale based on {\it Hipparcos} parallaxes for subdwarfs
(from Carretta et al. (2000) who estimate the total uncertainty at $\pm0.12$
mag).  Re-analysing those data using equation 1 with $\rho=-3.51\pm0.20$ 
we find a PL($K$) relation
zero-point of $\delta=-7.34\pm0.04$ (internal error), or
$\delta=-7.34\pm0.13$ allowing for the uncertainty in the cluster scale. The
points for the individual Miras are illustrated in Fig.~\ref{fig_pl}.

\subsection{The Galactic Bulge}
 Glass et al. (1995) discussed long period variables in the Sgr~I window of
the Galactic Bulge and derived a similar slope for the PL($K$) relation to
the one found here (section 2) for the LMC. Re-analysing these, as above,
gives $\delta+(m-M)_{GC}=7.40\pm0.05$. Using the Eisenhauer et al. (2005)
distance to the centre, with the relativistic correction suggested by Zucker
et al. (2006), gives $(m-M)_{GC}= 14.44\pm0.09$, and provides a zero-point
of $\delta=-7.04\pm0.11$ for these Galactic Bulge Miras. The similarity of
this to the other values found here suggests that any abundance effects on
the PL($K$) relation zero-point must be small.

In the conclusion below we do not average this Bulge distance along with the
other values to get a mean zero-point for the Galaxy. If we did so the
agreement of the Galactic mean with the LMC would be even closer. However,
there are systematic uncertainties associated with the Bulge (including its
shape and structure) that suggest it should not be treated in the same way.

\section{Conclusions}
In Paper~I (section 2.2) we noted a caveat with regard to temporal changes
in the light distribution across the stellar disk as problematic for the
interpretation of the parallaxes. This is particularly so because the
angular diameters of the Miras are two or three times larger than their
parallaxes. More recent work (e.g. Ragland et al. 2006; Woodruff et al.
2008) confirms that the diameters are large, variable and non-uniform. These
same references further highlight the disagreement between observation and
theory, emphasizing our very limited understanding of the atmospheres of
these stars or their variations. This is clearly a real problem, but the
agreement on the distance scale achieved by the different methods summarized
above may indicate that the net effect is small.
 
 As noted, the various results discussed above are in good agreement with
each other. UX Cyg, is brighter than we would expect from the PL($K$)
relation and this may be because of hot bottom burning. We take a simple
mean of solutions 10, 25, 26 from Table~\ref{zp}, to set a mean value of
$\delta =
-7.25 \pm 0.07$ for the O-rich Mira variables in the Galaxy. This is in
reasonable agreement with the LMC value of $\delta = -7.15 \pm 0.06$.

Solution 24, for the C-rich Miras, of $\delta = -7.18 \pm 0.37$ is in
agreement with the above result for O-rich Miras and with the LMC value for
C-stars of $\delta = -7.24 \pm 0.07$ This supports an identical PL($K$)
relation for O- and C-rich variables.

The O-rich result represents the best value for large amplitude variables,
but it should be used with caution on stars with periods in excess of 400
days. The $\rm H_2O$ parallaxes offer the most likely significant
improvement in this result in the near future.

\section*{Acknowledgments} We are grateful to Fred Marang and Francois van
Wyk for making the new infrared observation reported in the appendix. 

This publication makes use of data products from the Two Micron All Sky
Survey (2MASS), which is a joint project of the University of Massachusetts
and the Infrared Processing and Analysis Center/California Institute of
Technology, funded by the National Aeronautics and Space Administration and
the National Science Foundation. 

This paper utilizes public domain data obtained by the MACHO Project,
jointly funded by the US Department of Energy through the University of
California, Lawrence Livermore National Laboratory under contract No.
W-7405-Eng-48, by the National Science Foundation through the Center for
Particle Astrophysics of the University of California under cooperative
agreement AST-8809616, and by the Mount Stromlo and Siding Spring
Observatory, part of the Australian National University.

\section*{APPENDIX: Infrared Photometry and Basic Data}

The infrared $K$ magnitudes used here are largely those used in Paper~I, with
updates where significant quantities of new data are available.
New infrared photometry has been obtained from SAAO for 10 stars: W Psc, WW
Vel, RT Crt, S CrB, S Ser, BG Ser, R Ser, RU Her, T Her and W Sex. This is
listed in Table~\ref{ir} (the full table is available on-line) together
with previously published data for the same stars. Six of these stars have
sufficient data to derive periods (more than 9 observations) and their light
curves are illustrated in Fig.~\ref{fig-lc}. A description of the
photometer, telescope etc. was given in Whitelock et al. (2000).
 
\subsection{Interstellar Extinction}
The interstellar extinction corrections used here are derived in a different
way from those used in Paper~I. A first estimate is made of the distance to
the Mira assuming no interstellar extinction and equation 1 of section 2
with $\rho=-3.51\pm0.20$ and $\delta=-7.25$. The extinction is then
estimated using the Drimmel et al. (2003) three dimensional Galactic
extinction model, including the rescaling factors that correct the dust
column density to account for small scale structure seen in the DIRBE data,
but not described explicitly by the model. The measured mean $K$ mag is then
corrected for extinction following the reddening law given by Glass (1999)
and the procedure iterated; two iterations usually suffice.

This procedure gives significantly different, in many cases larger,
extinctions for individual stars from the statistical method applied by
Whitelock, Marang \& Feast (2000) and used in Paper~I. The biggest
difference among the O-rich stars is for UX Oph which has $A_V\sim 1.51$ mag
according to the method used here, in contrast to the $A_V\sim 0.23$ mag
used in Paper~I. Nevertheless, the net effect on the derived zero-point is
negligible, first because the reddening at $K$ is low, and secondly, because
the stars with the highest weight are the closest and have the least
reddening.

\subsection{Pulsation periods and amplitudes}
 Pulsation periods, $P_K$, and peak-to-peak amplitudes, $\Delta K$, were
derived for the stars with 9 or more SAAO observations. These are listed in
Table~\ref{new} along with Fourier mean magnitudes, $JHKL$, for all of the
stars with new observations (see also Fig.~\ref{fig-lc}).

 In general we use the pulsation periods tabulated in the General Catalogue
of Variable Stars (GCVS, Kholopov et al. 1985) for the parallax analysis, as
these are usually based on large data sets. However, Whitelock et al. (2000)
noted the very significant difference between the GCVS and {\it Hipparcos}
periods for RT Crt, WW Vel and W Sex (their fig.~1). We now have sufficient
infrared data to be confident that the {\it Hipparcos} periods are correct
for RT Crt and WW Vel. W Sex is a low amplitude variable C-star and its
period is not well determined from the IR observations although the {\it
Hipparcos} value of 200 d seems better than the 134 d GCVS value. We
therefore use the {\it Hipparcos} values for all three stars in the parallax 
analysis, in preference to the GCVS values (see also Fig.~\ref{fig-lc}).

Note that although the period derived from the infrared data, $P_K$, for
S~Ser, 230 days, is significantly different from the 372 days given in the
GCVS ({\it Hipparcos} derived 376), we use the GCVS value in the analysis as
there are insufficient IR data to be confident in any period derived from
them. Note also that although BG~Ser is classified as a Mira, its amplitude,
$\Delta K=0.31$ is low and more typical of a semi-regular variable. The
values used in the analysis are listed in Table~\ref{dist}.

\subsection{The Data}

Table~\ref{dist} contains the material used in, and derived from, 
the analysis
related to the infrared magnitudes. The columns are as follows: \\
(1) the GCVS variable star name;\\ 
(2) the {\it Hipparcos} catalogue number; \\
(3) the variability type: Mostly Mira (M) or semi-regular (SR) variables,
with two slow irregulars (L or LB) among the C-stars; \\
(4) P, the period used in the analysis, which is generally that 
tabulated in the GCVS (see section 6.2 for RT~Crt, WW~Vel and W~Sex); \\
(5) $\Delta H_p$, the peak-to-peak amplitude of the
variations in the {\it Hipparcos} mag; stars with large amplitudes, i.e., 
$\Delta H_p> 1.5 $ mag (O-rich) or $\Delta H_p> 1.0 $ mag (C-rich), 
are considered to be very similar to the Miras for most purposes; \\
(6) $K$, the mean $K$ mag, {\bf before} correcting for interstellar
extinction;\\ 
(7) the number of observations used to derive the mean $K$ in column (6) 
- where there are 9 or more observations the value will probably be 
close to the true mean (see below for more details of the non-SAAO photometry); \\
(8) $A_V$, the interstellar extinction at $V$ (see above), but note 
that this depends on the distance and the reddening should be regarded as 
a lower limit if it is uncertain that the star lies on 
the Mira PL($K$) relation, i.e. if there is no entry in the ``dist" column 
(because almost all of the non-Mira variables lie on PL($K$) relations 
above the Mira one);\\ 
(9) dist, the distance in kpc derived from equation 1, with
 $\rho=-3.51\pm0.20$ and 
$\delta=-7.25$ (the identical relation was used for O- and C-rich
stars), dist is not given for stars which do not, or may not, lie
on the Mira PL($K$) relation, i.e. low amplitude variables.\\

Where there were more than 5 observations contributing to the mean $K$ mag
listed by Whitelock et al. (2000), or new SAAO photometry,
these values were used. Otherwise $K$ mags were taken from the literature
and particularly from the compilation by Gezari, Pitts \& Schmitz (1999).
The following references which were not cited by Gezari et al., or which are
more recent, were also used: 2MASS (Cutri et al. 2003), DIRBE (Smith, Price
\& Baker 2004), Noguchi et al. (1981), Taranova \& Shenavrin (2004), Chen et
al. (1984), Gao, Chen \& Zhang (1985), Kerschbaum, Lebzelter \& Lazaro
(2001), Sun \& Zhang (1998), Chen et al. (1988)\footnote{The Chen et al.
value for R~Lyn ($K=0.52$) was not used as it is inconsistent with other
measurements.}, Le Bertre et al. (2003).

Where necessary and practical observations from the literature were
transformed to the SAAO system (e.g. Carter 1990, Carpenter 2001). Most of
the stars discussed here were sufficiently bright to be saturated in 2MASS
and are quoted in the catalogue (Cutri et al. 2003) with large errors, these
values were only used where little else was available. While there should be
no problem with saturation of the DIRBE photometry (Smith et al. 2004),
there is some risk of confusion with other sources in the large aperture.
More importantly there is very little information available on the
calibration or transformation for the DIRBE system; following Beverly Smith
(2007 private communication) we assume that a $K=0$ mag star has a flux of
630 Jy. Although the DIRBE photometry was integrated over a period of days
or weeks, a DIRBE $K$ mag was only given the same weight as any other
measurement when the mean was established for Table~\ref{dist}.

\renewcommand{\thetable}{A\arabic{table}}
\setcounter{table}{0}
\begin{table}
\centering
\caption{New infrared photometry from SAAO (the full table is available on-line).}\label{ir}
\begin{center}
\begin{tabular}{rrrrr}
\hline
\multicolumn{1}{c}{JD--2440000}& \multicolumn{1}{c}{$J$}& \multicolumn{1}{c}{$H$}&
\multicolumn{1}{c}{$K$}& \multicolumn{1}{c}{$L$}\\
\multicolumn{1}{c}{(day)}& \multicolumn{4}{c}{(mag)}\\
\hline
\multicolumn{5}{l}{ W Psc Hip 4652}\\
  9356.26 &  6.91 &  6.01 &  5.69 &  5.34\\
 10053.29 &  6.96 &  6.02 &  5.69 &  5.24\\
 10295.64 &  6.92 &  6.01 &  5.65 &  5.26\\
 10320.57 &  7.22 &  6.33 &  5.93 &  5.44\\
 10360.46 &  7.44 &  6.57 &  6.14 &  5.80\\
 10414.31 &  7.28 &  6.41 &  6.08 &  5.52\\
 10721.47 &  7.34 &  6.47 &  6.05 &  5.69\\
 10796.28 &  7.13 &  6.21 &  5.89 &  5.38\\
 11068.27 &  7.13 &  6.22 &  5.83 &  5.41\\
 11104.27 &  7.47 &  6.58 &  6.16 &  5.72\\
 12977.29 &  7.55 &  6.66 &  6.24 &  5.77\\
\hline
\end{tabular}
\end{center}
\end{table}

\begin{table*}
\centering
\caption{Fourier mean photometry and other data for stars with new SAAO 
$JHKL$ observations}\label{new}
\begin{center}
\begin{tabular}{rrcccrrcccr}
\hline
\multicolumn{1}{c}{Hip} & \multicolumn{1}{c}{name} &  $J$ & $H$ & $K$ 
& \multicolumn{1}{c}{$L$} &  $\Delta K$ & \multicolumn{1}{c}{no.}& 
$P$ & $P_K$ & typ \\
\multicolumn{1}{c}{No.} & & \multicolumn{5}{c}{(mag)} & & 
\multicolumn{2}{c}{(d)} & \\
\hline
  4652 &  W Psc  &      7.18 &  6.28 &  5.92 &   5.47 & 0.62 & 11 &  188 & 187  & M  \\
 48316 &  W Sex  &      5.09 &  3.94 &  3.48 &   3.05 & 0.11 & 21 &  200 & 126  & SR \\
 52988 &  WW Vel &      3.72 &  2.70 &  2.23 &   1.71 & 0.69 & 38 &  392 & 397  & M  \\
 53915 &  RT Crt &      5.88 &  5.04 &  4.58 &   3.96 & 0.88 & 16 &  183 & 185  & M  \\
 75143 &  S CrB  &      1.69 &  0.74 &  0.22 & --0.32 & 0.64 &  8 &  360 & -    & M  \\
 75170 &  S Ser  &      3.45 &  2.42 &  1.88 &   1.31 & 0.41 & 10 &  372 & 230  & M  \\
 77027 &  BG Ser &      1.95 &  0.85 &  0.39 & --0.04 & 0.31 &  8 &  404 & -    & M  \\
 77615 &  R Ser  &      2.03 &  1.11 &  0.63 &   0.03 & 0.70 & 13 &  356 & 355  & M  \\
 79233 &  RU Her &      1.96 &  0.88 &  0.34 & --0.25 & 0.73 &  9 &  485 & 482  & M  \\
 88923 &  T Her  &      4.46 &  3.67 &  3.35 &   2.91 & 0.42 &  4 &  165 & -    & M  \\
\hline
\end{tabular}
\end{center}
\end{table*}

\renewcommand{\thefigure}{A\arabic{figure}}
\setcounter{figure}{0}
\begin{figure}
\includegraphics[width=\columnwidth,clip=true]{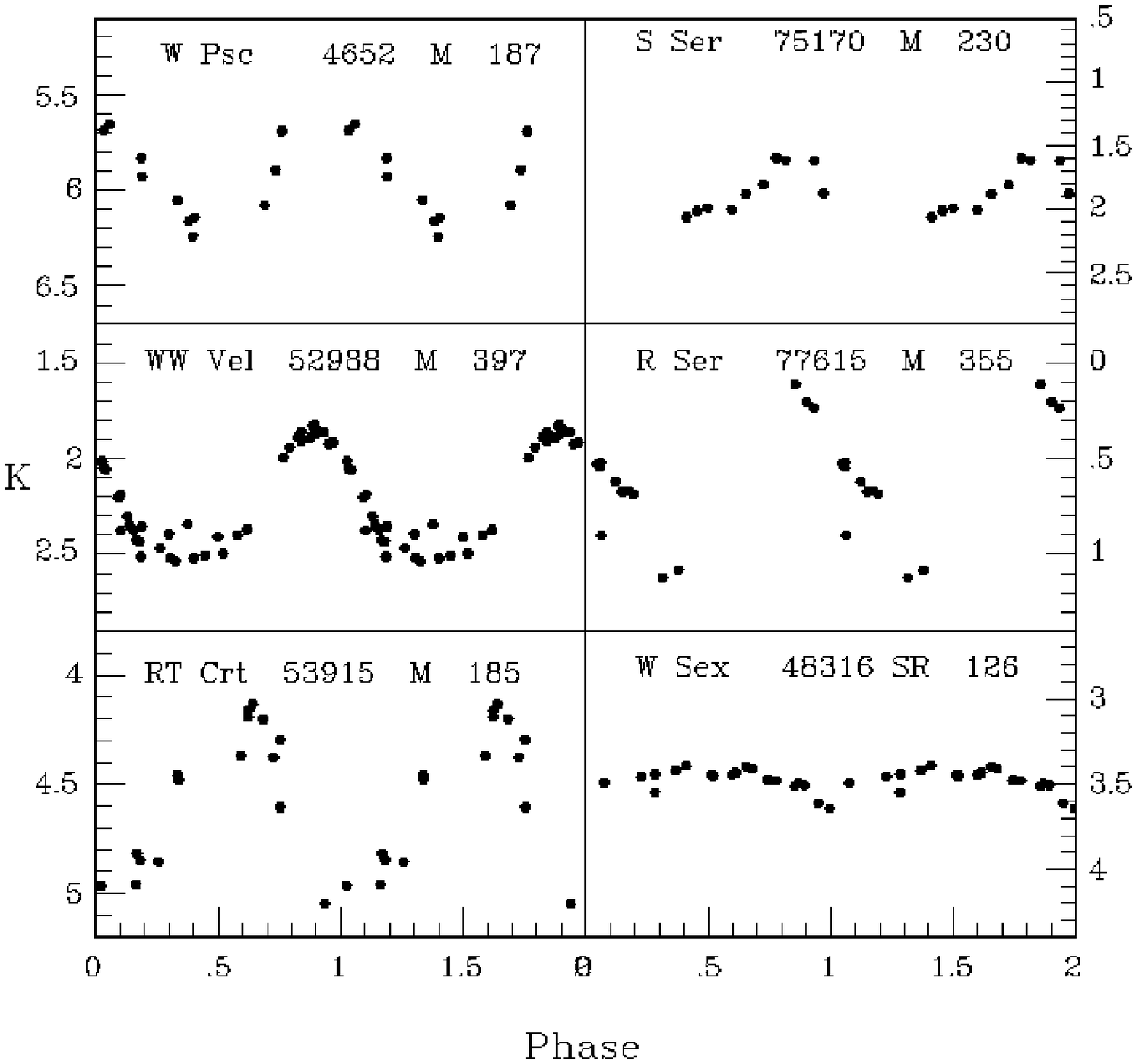}
\caption{Light-curves for Mira variables with new photometry and
observations on at least 10 epochs. These are plotted at arbitrary phase and
each point is shown twice to emphasize the periodicity. The name, {\it
Hipparcos} number, type of variability (M or SR) and period at which
the data are phased, is given on each plot.}\label{fig-lc}
\end{figure}

\begin{center}
\onecolumn
\begin{longtable}{lrcccrrcc}
\caption[Infrared and other data for variable stars.]{Infrared and other data 
for variable stars.}\label{dist} \\
\hline
\multicolumn{1}{c}{Name} & \multicolumn{1}{c}{Hip} &
\multicolumn{1}{c}{type} & \multicolumn{1}{c}{P} & 
\multicolumn{1}{c}{$\Delta H_P$} & \multicolumn{1}{c}{$K$} &
\multicolumn{1}{c}{no.}&
\multicolumn{1}{c}{$A_V$} & dist\\
& \multicolumn{1}{c}{ no.} & & \multicolumn{1}{c}{(d)}&  
\multicolumn{2}{c}{(mag)} & & 
\multicolumn{1}{c}{(mag)} & \multicolumn{1}{c}{(kpc)}\\
\hline
\endfirsthead
\hline
\multicolumn{1}{c}{Name} & \multicolumn{1}{c}{Hip} &
\multicolumn{1}{c}{type} & \multicolumn{1}{c}{P} & 
\multicolumn{1}{c}{$\Delta H_P$} & \multicolumn{1}{c}{$K$} &
\multicolumn{1}{c}{no.}&
\multicolumn{1}{c}{$A_V$} & dist\\
& \multicolumn{1}{c}{ no.}& & \multicolumn{1}{c}{(d)}&  \multicolumn{2}{c}{(mag)} & & 
\multicolumn{1}{c}{(mag)} & \multicolumn{1}{c}{(kpc)}\\
\hline
\endhead
 \multicolumn{9}{l}{{Continued on next page\ldots}} \\
\endfoot

  \\ \hline
\endlastfoot
\multicolumn{9}{l}{\bf Data for O-rich variables} \\
Z Peg   &      8 &M  & 335 & 4.01 &  1.09 &   7 & 0.11 &  0.59 \\
SV And  &    344 &M  & 316 & 3.96 &  2.45 &   4 & 0.26 &  1.05 \\
SW Scl  &    516 &SR & 146 & 1.30 &  3.72 &   4 & 0.04 &  \\
RU Oct  &    703 &M  & 373 & 2.23 &  2.68 &  14 & 0.26 &  1.30 \\
SS Cas  &    781 &M  & 141 & 3.00 &  3.48 &   2 & 0.41 &  0.95 \\
S Scl   &   1236 &M  & 362 & 4.50 &  0.31 & 122 & 0.04 &  0.43 \\
T Cas   &   1834 &M  & 445 & 2.16 & --1.04 &   8 & 0.15 & 0.27 \\
R And   &   1901 &M  & 409 & 4.27 &  0.01 &   7 & 0.17 &  0.41 \\
T Scl   &   2286 &M  & 202 & 2.88 &  3.74 &   9 & 0.06 &  1.39 \\
TU And  &   2546 &M  & 317 & 3.52 &  2.11 &   5 & 0.11 &  0.90 \\
W Psc   &   4652 &M  & 188 & 2.61 &  5.92 &  11 & 0.17 &  3.60 \\
U Per   &   9306 &M  & 320 & 2.36 &  0.89 &   5 & 0.27 &  0.51 \\
Y Eri   &   9767 &M  & 303 & 2.10 &  1.79 &  14 & 0.07 &  0.75 \\
R Ari   &  10576 &M  & 187 & 3.61 &  3.94 &  12 & 0.20 &  1.44 \\
W And   &  10687 &M  & 396 & 5.10 &  0.43 &   5 & 0.11 &  0.49 \\
$o$ Cet &  10826 &M  & 332 & 4.39 & --2.45 & 104 & 0.04 & 0.11 \\
R Cet   &  11350 &M  & 166 & 4.25 &  2.54 &  30 & 0.10 &  0.70 \\
S Tri   &  11423 &M  & 242 & 1.46 &  3.86 &   3 & 0.21 &  1.66 \\
U Cet   &  11910 &M  & 235 & 3.88 &  2.77 &  25 & 0.06 &  0.99 \\
R Tri   &  12193 &M  & 267 & 3.96 &  0.97 &  13 & 0.13 &  0.47 \\
T Ari   &  13092 &SR & 317 & 1.66 &  0.17 &  23 & 0.23 &  0.37 \\
R Hor   &  13502 &M  & 408 & 4.48 & --0.93 &  52 & 0.07 & 0.27 \\
T Hor   &  14042 &M  & 218 & 3.29 &  3.32 &  14 & 0.06 &  1.21 \\
X Cet   &  15465 &M  & 177 & 2.68 &  4.23 &  15 & 0.16 &  1.59 \\
RT Eri  &  16647 &M  & 371 & 2.60 &  0.39 &   5 & 0.28 &  0.45 \\
T Eri   &  18336 &M  & 252 & 3.73 &  2.42 &  23 & 0.09 &  0.89 \\
W Eri   &  19567 &M  & 377 & 4.04 &  1.51 &  16 & 0.16 &  0.77 \\
RS Eri  &  20045 &M  & 296 & 3.05 &  1.22 &   5 & 0.12 &  0.57 \\
R Ret   &  21252 &M  & 278 & 3.42 &  1.74 &  22 & 0.09 &  0.69 \\
RX Tau  &  21600 &M  & 332 & 2.12 &  1.21 &  10 & 0.39 &  0.61 \\
R Cae   &  21766 &M  & 391 & 3.56 &  0.59 & 116 & 0.05 &  0.52 \\
X Cam   &  22127 &M  & 144 & 3.69 &  3.64 &   1 & 0.28 &  1.04 \\
SU Dor  &  22256 &M  & 236 & 3.44 &  5.09 &  11 & 0.03 &  2.90 \\
T Lep   &  23636 &M  & 368 & 2.89 &  0.09 &  16 & 0.06 &  0.39 \\
U Dor   &  24055 &M  & 394 & 3.38 &  1.17 &  35 & 0.11 &  0.68 \\
S Pic   &  24126 &M  & 428 & 4.37 &  0.72 &  93 & 0.05 &  0.59 \\
T Pic   &  24468 &M  & 201 & 4.04 &  4.26 &  14 & 0.07 &  1.77 \\
R Aur   &  24645 &M  & 458 & 3.72 & --0.69 &   4 & 0.16 & 0.32 \\
T Col   &  24824 &M  & 226 & 2.79 &  1.96 &  14 & 0.06 &  0.66 \\
R Oct   &  25412 &M  & 405 & 4.00 &  0.75 &   6 & 0.38 &  0.57 \\
S Ori   &  25673 &M  & 414 & 3.05 & --0.07 & 101 & 0.73 & 0.39 \\
S Col   &  27286 &M  & 326 & 3.22 &  1.59 &  12 & 0.07 &  0.73 \\
U Ori   &  28041 &M  & 368 & 4.90 & --0.75 &  56 & 0.14 & 0.27 \\
RS Aur  &  28714 &SR & 171 & 0.73 &  3.10 &   2 & 0.33 &  \\
X Aur   &  29441 &M  & 164 & 2.86 &  3.43 &   3 & 0.32 &  1.03 \\
V Mon   &  30326 &M  & 340 & 4.20 &  1.08 &  20 & 0.29 &  0.59 \\
RV Pup  &  32115 &M  & 188 & 2.96 &  3.62 &  17 & 0.30 &  1.25 \\
X Gem   &  32512 &M  & 264 & 3.10 &  1.86 &   7 & 0.05 &  0.71 \\
X Mon   &  33441 &SR & 156 & 1.17 &  2.87 &  20 & 0.39 &  \\
R Lyn   &  33824 &M  & 379 & 4.29 &  2.14 &   4 & 0.14 &  1.04 \\
R Gem   &  34356 &M  & 370 & 4.31 &  1.67 &   4 & 0.04 &  0.82 \\
L2 Pup  &  34922 &SR & 141 & 0.71 & --2.24 &  71 & 0.04 & \\
V Gem   &  35812 &M  & 275 & 3.31 &  2.81 &   9 & 0.14 &  1.12 \\
TT Mon  &  36043 &M  & 323 & 3.86 &  1.95 &   7 & 0.38 &  0.84 \\
VX Aur  &  36314 &M  & 322 & 3.07 &  1.53 &   3 & 0.14 &  0.70 \\
RX Mon  &  36394 &M  & 346 & 3.72 &  2.67 &   6 & 0.11 &  1.24 \\
S Vol   &  36423 &M  & 395 & 3.67 &  2.41 &   3 & 0.59 &  1.18 \\
Z Pup   &  36669 &M  & 509 & 3.89 &  1.33 &  72 & 0.44 &  0.87 \\
S CMi   &  36675 &M  & 333 & 3.85 &  0.48 &  11 & 0.05 &  0.44 \\
U CMi   &  37459 &M  & 413 & 3.41 &  2.61 &   9 & 0.32 &  1.36 \\
W Pup   &  37893 &M  & 120 & 3.28 &  3.54 &  20 & 0.44 &  0.87 \\
SU Pup  &  38772 &M  & 340 & 5.45 &  2.58 &  11 & 0.34 &  1.16 \\
AS Pup  &  39967 &M  & 325 & 2.59 &  0.27 &   6 & 0.69 &  0.38 \\
R Cnc   &  40534 &M  & 362 & 3.32 & --0.62 &  19 & 0.07 & 0.28 \\
SV Pup  &  40593 &M  & 167 & 3.78 &  3.59 &   4 & 0.24 &  1.13 \\
S Hya   &  43653 &M  & 257 & 4.08 &  2.85 &  11 & 0.16 &  1.09 \\
T Hya   &  43835 &M  & 299 & 3.98 &  2.36 &  14 & 0.09 &  0.97 \\
W Cnc   &  44995 &M  & 393 & 3.90 &  1.04 &   5 & 0.10 &  0.64 \\
R Car   &  46806 &M  & 309 & 4.27 & --1.35 &  74 & 0.10 & 0.18 \\
X Hya   &  47066 &M  & 301 & 3.73 &  0.65 &  13 & 0.18 &  0.44 \\
R LMi   &  47886 &M  & 372 & 3.58 & --0.34 &  10 & 0.04 & 0.33 \\
R Leo   &  48036 &M  & 310 & 3.46 & --2.55 &  53 & 0.03 & 0.10 \\
S LMi   &  48520 &M  & 234 & 4.37 &  3.81 &   3 & 0.04 &  1.60 \\
V Leo   &  49026 &M  & 273 & 3.90 &  3.17 &   6 & 0.06 &  1.32 \\
X Ant   &  49524 &M  & 162 & 3.41 &  5.41 &   6 & 0.20 &  2.56 \\
S Car   &  49751 &M  & 149 & 2.79 &  1.87 &  34 & 0.18 &  0.47 \\
W Vel   &  50230 &M  & 395 & 3.62 &  0.56 &  13 & 0.42 &  0.51 \\
V Ant   &  50697 &M  & 303 & 3.97 &  2.10 &   9 & 0.30 &  0.86 \\
S Sex   &  51791 &M  & 265 & 2.72 &  3.42 &  12 & 0.21 &  1.44 \\
R UMa   &  52546 &M  & 302 & 4.18 &  1.37 &  13 & 0.07 &  0.62 \\
WX Vel  &  52887 &M  & 412 & 3.72 &  1.81 &   8 & 0.66 &  0.92 \\
WW Vel  &  52988 &M  & 392 & 2.77 &  2.23 &  38 & 0.58 &  1.08 \\
CI Vel  &  53853 &M  & 143 & 2.38 &  6.38 &   7 & 0.89 &  3.58 \\
RT Crt  &  53915 &M  & 183 & 3.20 &  4.58 &  16 & 0.12 &  1.91 \\
DN Hya  &  57009 &M  & 182 & 3.20 &  5.37 &   5 & 0.24 &  2.73 \\
X Cen   &  57642 &M  & 315 & 3.60 &  1.10 &  14 & 0.32 &  0.56 \\
W Cen   &  58107 &M  & 202 & 3.45 &  1.95 &  17 & 0.44 &  0.60 \\
R Com   &  58854 &M  & 363 & 4.35 &  2.21 &   8 & 0.06 &  1.04 \\
R Crv   &  60106 &M  & 317 & 4.10 &  1.88 &  19 & 0.17 &  0.81 \\
XZ Cen  &  60502 &M  & 291 & 1.95 &  1.51 &   4 & 0.23 &  0.64 \\
T CVn   &  61009 &M  & 290 & 1.41 &  2.04 &   6 & 0.04 &  0.82 \\
U Cen   &  61286 &M  & 220 & 3.53 &  3.13 &   8 & 0.61 &  1.09 \\
T UMa   &  61532 &M  & 257 & 4.13 &  2.74 &   6 & 0.03 &  1.04 \\
R Vir   &  61667 &M  & 146 & 3.27 &  2.05 &  10 & 0.06 &  0.51 \\
S UMa   &  62126 &M  & 226 & 3.00 &  3.04 &   3 & 0.04 &  1.09 \\
U Vir   &  62712 &M  & 207 & 3.80 &  4.01 &   7 & 0.07 &  1.61 \\
BZ Vir  &  63501 &M  & 151 & 2.15 &  5.35 &   4 & 0.24 &  2.37 \\
V CVn   &  65006 &SR & 192 & 1.13 &  1.12 &   6 & 0.10 &  \\
R Hya   &  65835 &M  & 389 & 2.71 & --2.47 &  53 & 0.12 & 0.13 \\
S Vir   &  66100 &M  & 375 & 4.25 &  0.33 &  13 & 0.10 &  0.45 \\
T Cen   &  66825 &SR &  90 & 1.79 &  2.50 &  27 & 0.12 &  0.45 \\
RT Cen  &  67359 &M  & 255 & 1.40 &  2.69 &  18 & 0.23 &  1.01 \\
R CVn   &  67410 &M  & 329 & 3.36 &  0.45 &   5 & 0.03 &  0.43 \\
W Hya   &  67419 &SR & 361 & 2.02 & --3.16 &  24 & 0.05 & 0.09 \\
RX Cen  &  67626 &M  & 328 & 4.99 &  2.77 &  16 & 0.25 &  1.24 \\
RU Hya  &  69346 &M  & 332 & 3.81 &  1.57 &  15 & 0.16 &  0.72 \\
R Cen   &  69754 &M  & 546 & 2.63 & --0.70 &  66 & 0.21 & 0.36 \\
U UMi   &  69816 &M  & 331 & 2.51 &  0.71 &   2 & 0.06 &  0.49 \\
S Boo   &  70291 &M  & 271 & 3.89 &  3.11 &   2 & 0.02 &  1.28 \\
RW Lup  &  70590 &M  & 331 & 1.07 &  2.52 &   9 & 0.44 &  1.11 \\
RS Vir  &  70669 &M  & 354 & 3.71 &  1.10 &  30 & 0.09 &  0.61 \\
R Boo   &  71490 &M  & 223 & 3.82 &  2.11 &   6 & 0.06 &  0.71 \\
RR Boo  &  72300 &M  & 195 & 3.70 &  5.19 &   2 & 0.02 &  2.66 \\
Y Lib   &  74350 &M  & 276 & 4.17 &  3.29 &  15 & 0.24 &  1.40 \\
RT Boo  &  74802 &M  & 274 & 2.04 &  2.56 &   3 & 0.06 &  1.00 \\
S CrB   &  75143 &M  & 360 & 4.09 &  0.32 &  22 & 0.05 &  0.43 \\
S Lib   &  75144 &M  & 193 & 2.32 &  4.51 &  14 & 0.48 &  1.90 \\
S Ser   &  75170 &M  & 372 & 2.69 &  1.88 &  10 & 0.13 &  0.91 \\
RS Lib  &  75393 &M  & 218 & 2.58 & --0.07 &  35 & 0.26 & 0.25 \\
S UMi   &  75847 &M  & 331 & 2.61 &  3.04 &   3 & 0.08 &  1.43 \\
R Nor   &  76377 &M  & 508 & 4.53 &  1.27 &  55 & 0.59 &  0.84 \\
BG Ser  &  77027 &M  & 404 & 2.04 &  0.39 &   8 & 0.36 &  0.48 \\
T Nor   &  77058 &M  & 241 & 3.93 &  2.16 &   6 & 0.49 &  0.75 \\
X CrB   &  77460 &M  & 241 & 3.59 &  3.42 &   2 & 0.06 &  1.36 \\
R Ser   &  77615 &M  & 356 & 4.54 &  0.63 &  13 & 0.12 &  0.49 \\
RZ Sco  &  78746 &M  & 157 & 1.93 &  4.19 &  15 & 0.60 &  1.41 \\
Z Sco   &  78872 &M  & 343 & 1.78 &  1.48 &  15 & 0.82 &  0.69 \\
RU Her  &  79233 &M  & 485 & 4.46 &  0.36 &  20 & 0.19 &  0.54 \\
U Her   &  80488 &M  & 406 & 3.88 & --0.29 &  20 & 0.14 & 0.36 \\
R Dra   &  81014 &M  & 246 & 3.85 &  2.26 &   4 & 0.08 &  0.81 \\
SS Her  &  81026 &M  & 107 & 2.68 &  5.08 &   8 & 0.24 &  1.64 \\
AS Her  &  81506 &M  & 269 & 3.44 &  1.94 &   5 & 0.16 &  0.74 \\
S Her   &  82516 &M  & 307 & 4.39 &  1.30 &   5 & 0.20 &  0.60 \\
Z Ara   &  82695 &M  & 289 & 3.46 &  3.20 &   4 & 0.59 &  1.37 \\
RS Sco  &  82833 &M  & 320 & 4.45 &  0.40 &  13 & 0.25 &  0.41 \\
RR Sco  &  82912 &M  & 281 & 3.34 & --0.23 &  54 & 0.22 & 0.28 \\
SY Her  &  83304 &M  & 117 & 3.04 &  4.33 &   7 & 0.16 &  1.24 \\
UX Oph  &  83582 &M  & 117 & 3.15 &  4.71 &   3 & 1.51 &  1.40 \\
Z Oph   &  84763 &M  & 349 & 3.80 &  4.01 &   5 & 0.57 &  2.27 \\
RS Her  &  84948 &M  & 220 & 3.25 &  2.92 &   4 & 0.16 &  1.01 \\
Z Oct   &  86836 &M  & 335 & 2.48 &  2.72 &   2 & 0.45 &  1.22 \\
T Her   &  88923 &M  & 165 & 3.74 &  3.36 &   8 & 0.19 &  1.01 \\
R Pav   &  89258 &M  & 229 & 3.65 &  2.90 &  12 & 0.25 &  1.02 \\
W Lyr   &  89419 &M  & 198 & 3.37 &  3.22 &   5 & 0.10 &  1.08 \\
RY Oph  &  89568 &M  & 150 & 3.00 &  2.96 &   8 & 0.51 &  0.77 \\
AL Dra  &  90474 &M  & 330 & 3.20 &  2.99 &   4 & 0.23 &  1.38 \\
RV Sgr  &  90493 &M  & 316 & 4.44 &  1.62 &  18 & 0.35 &  0.71 \\
RS Dra  &  91316 &SR & 283 & 0.61 &  1.96 &   3 & 0.27 &  \\
R Aql   &  93820 &M  & 284 & 3.11 & --0.76 &  45 & 0.13 & 0.22 \\
SS Lyr  &  94438 &M  & 346 & 2.80 &  1.94 &   3 & 0.17 &  0.88 \\
RW Sgr  &  94489 &SR & 187 & 0.61 &  3.06 &   9 & 0.32 &  \\
R Sgr   &  94738 &M  & 270 & 4.28 &  2.06 &  13 & 0.28 &  0.78 \\
DD Cyg  &  96031 &M  & 148 & 2.90 &  6.29 &   1 & 0.61 &  3.55 \\
RT Aql  &  96580 &M  & 327 & 3.22 &  1.12 &   9 & 0.27 &  0.58 \\
BG Cyg  &  96647 &M  & 288 & 1.19 &  1.06 &   3 & 0.24 &  0.52 \\
RT Cyg  &  97068 &M  & 190 & 3.68 &  3.23 &   3 & 0.19 &  1.05 \\
chi Cyg &  97629 &M  & 408 & 5.51 & --1.91 &  11 & 0.10 & 0.17 \\
T Pav   &  97644 &M  & 244 & 4.31 &  2.87 &   9 & 0.31 &  1.06 \\
RR Sgr  &  98077 &M  & 336 & 4.56 &  0.69 &  14 & 0.28 &  0.49 \\
RU Sgr  &  98334 &M  & 240 & 3.93 &  2.18 &   5 & 0.24 &  0.76 \\
BQ Pav  &  98447 &M  & 110 & 3.01 &  6.58 &  16 & 0.17 &  3.35 \\
R Del   &  99802 &M  & 285 & 3.20 &  1.93 &   5 & 0.41 &  0.76 \\
RT Sgr  & 100033 &M  & 306 & 3.99 &  1.33 &   7 & 0.14 &  0.61 \\
CN Cyg  & 100048 &M  & 199 & 2.99 &  4.05 &   1 & 0.61 &  1.56 \\
V865 Aql& 100599 &M  & 367 & 3.20 &  1.60 &   7 & 0.42 &  0.78 \\
U Mic   & 101063 &M  & 334 & 3.55 &  1.84 &  21 & 0.15 &  0.83 \\
R Mic   & 101985 &M  & 139 & 3.58 &  3.66 &   7 & 0.18 &  1.03 \\
S Del   & 102246 &M  & 278 & 2.18 &  2.08 &   4 & 0.22 &  0.81 \\
V Aqr   & 102546 &SR & 244 & 1.13 &  0.58 &   3 & 0.23 &  \\
AM Cyg  & 102732 &M  & 371 & 2.42 &  1.88 &   2 & 0.44 &  0.89 \\
T Aqr   & 102829 &M  & 202 & 3.17 &  3.24 &  14 & 0.19 &  1.10 \\
RX Vul  & 103069 &M  & 457 & 1.54 &  1.06 &   2 & 0.34 &  0.71 \\
R Vul   & 104015 &M  & 164 & 3.60 &  3.24 &   2 & 0.44 &  0.94 \\
V Cap   & 104285 &M  & 276 & 3.51 &  3.48 &  11 & 0.15 &  1.53 \\
T Cep   & 104451 &M  & 388 & 3.45 & --1.76 &   6 & 0.33 & 0.17 \\
T Cap   & 105498 &M  & 269 & 3.20 &  3.24 &  18 & 0.22 &  1.35 \\
W Cyg   & 106642 &SR & 131 & 0.62 & --1.40 &   4 & 0.06 & \\
RU Cyg  & 107036 &SR & 233 & 0.75 & --0.29 &   3 & 0.14 & \\
TU Peg  & 107390 &M  & 321 & 2.97 &  1.10 &   4 & 0.30 &  0.57 \\
RS Peg  & 109610 &M  & 415 & 3.14 &  1.20 &   7 & 0.17 &  0.71 \\
DH Lac  & 109619 &M  & 289 & 2.83 &  4.56 &   1 & 0.43 &  2.58 \\
X Aqr   & 110146 &M  & 312 & 3.65 &  2.96 &  18 & 0.10 &  1.32 \\
UU Tuc  & 110451 &M  & 335 & 3.08 &  3.53 &  66 & 0.08 &  1.80 \\
RT Aqr  & 110509 &M  & 246 & 1.46 &  2.17 &  20 & 0.08 &  0.78 \\
T Gru   & 110697 &M  & 136 & 2.35 &  5.24 &  14 & 0.04 &  2.10 \\
S Gru   & 110736 &M  & 402 & 4.38 &  0.68 &  14 & 0.03 &  0.55 \\
S Lac   & 110972 &M  & 242 & 3.40 &  2.42 &   4 & 0.47 &  0.85 \\
SS Peg  & 111385 &M  & 425 & 3.94 &  1.15 &   6 & 0.12 &  0.71 \\
T Tuc   & 111946 &M  & 250 & 3.84 &  3.22 &   7 & 0.06 &  1.28 \\
SX Peg  & 112784 &M  & 304 & 3.53 &  3.15 &   7 & 0.22 &  1.41 \\
TV And  & 113405 &SR & 114 & 1.71 &  3.50 &   1 & 0.53 &  0.82 \\
RT Oct  & 113652 &M  & 180 & 4.22 &  5.72 &   3 & 0.43 &  3.16 \\
R Peg   & 114114 &M  & 378 & 4.11 &  0.52 &  10 & 0.20 &  0.49 \\
V Cas   & 114515 &M  & 229 & 3.17 &  0.86 &   4 & 0.22 &  0.40 \\
TY And  & 114757 &SR & 260 & 0.79 &  1.62 &   4 & 0.25 &  \\
W Peg   & 115188 &M  & 346 & 2.58 & --0.02 &   6 & 0.19 & 0.36 \\
S Peg   & 115242 &M  & 319 & 3.48 &  1.42 &  12 & 0.16 &  0.66 \\
R Aqr   & 117054 &M  & 387 & 3.82 & --1.02 & 104 & 0.06 & 0.25 \\
R Cas   & 118188 &M  & 430 & 4.24 & --1.79 &  13 & 0.10 & 0.19 \\

\multicolumn{9}{l}{\bf Data for C-rich variables} \\
  WZ Cas&     99 &SR & 186 & 0.44 &  0.61 &   8 & 0.17 &  \\
   W Cas&   4284 &M  & 399 & 2.02 &  2.77 &  14 & 0.64 & 1.44 \\
   R Scl&   6759 &SR & 375 & 1.13 & --0.09 & 157 & 0.07 &0.37 \\\
   X Cas&   9057 &M  & 420 & 1.44 &  2.37 &   4 & 0.59 & 1.24 \\
   R For&  11582 &M  & 385 & 3.30 &  1.21 & 100 & 0.04 & 0.68 \\
   Y Per&  16126 &M  & 238 & 0.61 &  3.64 &   1 & 0.48 & 1.50 \\
  SY Per&  19931 &SR & 474 & 1.11 &  1.82 &   5 & 0.51 & 1.05 \\
  AU Aur&  22796 &M  & 400 & 1.46 &  2.87 &   1 & 0.64 & 1.51 \\
   R Ori&  23165 &M  & 377 & 2.37 &  4.07 &  14 & 0.31 & 2.53 \\
   R Lep&  23203 &M  & 438 & 1.65 &  0.05 &  71 & 0.24 & 0.44 \\
   S Cam&  26753 &SR & 327 & 1.15 &  2.61 &   1 & 0.28 & 1.16 \\
   Y Tau&  27181 &SR & 242 & 0.22 &  0.26 &   6 & 0.17 &  \\
   V Aur&  30449 &M  & 349 & 1.72 &  2.99 &   1 & 0.33 & 1.45 \\
  CR Gem&  31349 &SR & 250 & 0.54 &  1.36 &   1 & 0.27 &  \\
   R CMi&  34474 &M  & 338 & 2.38 &  2.54 &  11 & 0.41 & 1.16 \\
  VX Gem&  34859 &M  & 391 & 1.87 &  3.13 &   9 & 0.03 & 1.68 \\
   T Lyn&  41058 &M  & 409 & 2.07 &  3.02 &  23 & 0.13 & 1.65 \\
   R Pyx&  42975 &M  & 369 & 1.70 &  2.50 &  11 & 0.36 & 1.20 \\
  UW Pyx&  43123 &LB & 423 & 1.50 &  2.62 &   5 & 0.65 & 1.40 \\
  IQ Hya&  45266 &SR & 397 & 1.77 &  2.84 &  15 & 0.52 & 1.49 \\
   W Sex&  48316 &SR & 200 & 0.54 &  3.47 &  20 & 0.15 &  \\
   V Hya&  53085 &L  & 532 & 2.15 & --0.71 &  75 & 0.13 &0.36 \\\
  BH Cru&  59844 &M  & 421 & 1.71 &  1.49 &  54 & 0.65 & 0.83 \\
  RU Vir&  62401 &M  & 444 & 1.73 &  1.79 &  46 & 0.08 & 0.99 \\
   V Cru&  63175 &M  & 380 & 1.84 &  2.80 &  96 & 0.94 & 1.41 \\
  RV Cen&  66466 &M  & 447 & 1.25 &  1.40 &  87 & 0.82 & 0.83 \\
   V CrB&  77501 &M  & 358 & 2.05 &  1.27 &  48 & 0.04 & 0.67 \\
  RR Her&  78721 &SR & 240 & 1.61 &  3.22 &   3 & 0.08 & 1.24 \\
   V Oph&  80550 &M  & 297 & 1.54 &  1.56 &  11 & 0.86 & 0.67 \\
  SZ Ara&  84059 &M  & 222 & 1.38 &  4.41 &   8 & 0.47 & 2.03 \\
  TT Cyg&  96836 &SR & 118 & 0.38 &  1.95 &   2 & 0.21 &  \\
  RS Cyg&  99653 &SR & 417 & 1.40 &  1.00 &   4 & 0.50 & 0.66 \\
  WX Cyg& 100113 &M  & 399 & 1.48 &  2.29 &   1 & 0.78 &  \\
   U Cyg& 100219 &M  & 460 & 2.18 &  1.11 &   7 & 0.59 & 0.74 \\
   V Cyg& 102082 &M  & 417 & 1.91 &  0.02 &   1 & 0.38 & 0.42 \\
  YY Cyg& 105539 &SR & 388 & 0.51 &  2.58 &   1 & 0.73 & \\
   S Cep& 106583 &M  & 486 & 1.79 & --0.10 &   1 & 0.38 &0.44 \\\
  RV Cyg& 107242 &SR & 263 & 0.25 &  0.37 &   2 & 0.29 &  \\
  RZ Peg& 109089 &M  & 439 & 3.12 &  2.48 &   2 & 0.44 & 1.35 \\
  ST And& 116681 &SR & 328 & 1.65 &  3.37 &  16 & 0.28 & 1.65 \\
\end{longtable}
\end{center}
\twocolumn
\end{document}